\documentclass[aps,notitlepage,twocolumn,nofootinbib]{revtex4-1}

\usepackage[brazil]{babel}
\usepackage{epsfig}
\usepackage{amssymb}
\usepackage{amsmath}
\usepackage{subeqnarray}
\usepackage{graphicx}
\usepackage{epstopdf}
\usepackage[colorlinks,citecolor=blue,linkcolor=blue,hyperindex]{hyperref}
\newcommand{\be}{\begin{equation}}
\newcommand{\ee}{\end{equation}}
\newcommand{\ben}{\begin{eqnarray}}
\newcommand{\een}{\end{eqnarray}}
 
\usepackage{pstricks}
\usepackage{color}

\begin{document}

\title{Finite-size effects on the Phase Structure of the Walecka Model}

\author{L. M. Abreu$^{1}$ {\footnote{email: luciano.abreu@ufba.br}} and
E. S. Nery$^{1}${\footnote{email: elenilsonnery@hotmail.com}}
}
\affiliation{$^{1}$Instituto de F{\'i}sica, Universidade Federal da
Bahia, 40210-340, Salvador-BA, Brazil}

\begin{abstract}

 In this work we investigate the finite-size effects on the phase structure of Walecka model within the framework of generalized Zeta-function, focusing on the influence of temperature as well as the number and length of compactified spatial dimensions. Here we concentrate on the situation of larger values of the coupling  between the scalar and fermion fields, in which a phase transition of first order takes place. The phase transitions are analyzed and compared with the system in the situations of one, two and three compactified spatial dimensions.  Our findings suggest that the thermodynamic behavior of the system depends on the length and number of spatial dimensions, with the symmetric phase being favored as the size of the system diminishes.

\end{abstract}
\maketitle

%%%%%%%%%%%%%%%%%%%%%%%%%%%%%%%%%%%%%%%%%%%%%%%%%%%%%%%%%%%%%%%%%%%%%%%%%%%%%%%%%%%%%%%%
%%%%%%%%%%%%%%%%%%%%%%%%%%%%%%%%%%%%%%%%%%%%%%%%%%%%%%%%%%%%%%%%%%%%%%%%%%%%%%%%%%%%%%%%
\section{Introduction}
%%%%%%%%%%%%%%%%%%%%%%%%%%%%%%%%%%%%%%%%%%%%%%%%%%%%%%%%%%%%%%%%%%%%%%%%%%%%%%%%%%%%%%%%
%%%%%%%%%%%%%%%%%%%%%%%%%%%%%%%%%%%%%%%%%%%%%%%%%%%%%%%%%%%%%%%%%%%%%%%%%%%%%%%%%%%%%%%%

One of the most interesting questions that has been receiving a great deal of attention in hadron and nuclear physics is the study of strongly interacting matter properties under the changes of the environment. As examples, we can highlight several phenomena: the phase diagram of nuclear and quark matter, relativistic degenerate gas phase transitions, quark-gluon plasma formation in heavy-ion collisions, phase structure of neutron stars and so on~\cite{Bellac,Kapusta}.

%A great deal of attention over recent decades has been devoted to the study of strongly interacting matter under extreme conditions.  In light of the theoretical grounds, effective quantum field theories at finite temperature have proved to be a very useful tool.

%The fundamental point of these interactions is their characteristic singular in low-energy, known as asymptotic freedom, which means treat %
%them by non-perturbative methods. Even in residual interactions between colorless hadrons.%
%Here the theories said effective in temperature and finite %
%density play an important role in the description of these systems. As for example in the %
%transition of phases, transport phenomena, nuclear matter and others \cite{Bellac,Kapusta}.%

In light of the theoretical grounds to treat these physical systems, effective quantum field theories of Quantum Chromodynamics (QCD) at finite temperature have proved to be very useful tools. 
In particular, one emblematic example is the Walecka model~\cite{Walecka:1974qa}. 
Different versions of this model have been largely employed as a laboratory to get insights on the thermodynamic behavior of hadronic matter, describing a reasonable number of phenomena in the sector of strong interactions~(for reviews see Refs.~\cite{JTheis:1983PRD,PhysRevC.58.1804,Saito,Bass:1998ca,Bender:2003jk,WEBER2005193,Li:2008gp,Delfino,Lavagno,Hayano,RocaMaza:2011qe,FUKUSHIMA201399,Casali,Dutra:2014qga,Mondal:2015tfa,Maslov:2015wba,Zhang:2017etr,Oertel:2016bki}).

A representative system which has been understood at least qualitatively via Walecka-like models is the nuclear matter, whose the exchange nucleon-nucleon interactions are discussed. Considering the scenario of finite temperature field theory, it can be thought as a gas of nucleons (associated with the Dirac field) embedded in a bath of scalar, vector and other types of particles constituting the hot and dense hadronic medium (associated with scalar, vectors and other fields). 
Therefore, interesting aspects of thermodynamic properties of this system can be investigated when it is under certain conditions, like finite temperature, finite chemical potential, external magnetic field, and others~\cite{Kapusta,Walecka:1974qa,JTheis:1983PRD,PhysRevC.58.1804,Saito,Bass:1998ca,Bender:2003jk,WEBER2005193,Li:2008gp,Delfino,Lavagno,Hayano,RocaMaza:2011qe,FUKUSHIMA201399,Casali,Dutra:2014qga,Mondal:2015tfa,Maslov:2015wba,Zhang:2017etr,Oertel:2016bki}.

Besides, there is a vast bibliography on the subject of the of finite-size effects on the thermodynamics of effective field theories for different physical phenomena.  Some interesting examples of these works are in Refs.~\cite{Abreu1,Ebert0,Ebert1,Hayashi,Abreu2,Braun,Palhares,Luciano1,Braun2,Flachi,Ebert2,Bhattacharyya1,Berg,Abreu3,Abreu6,Ebert3,Khanna,Bhattacharyya2,Bhattacharyya3,Abreu4,Abreu5,Abreu7,Bhattacharyya:2015kda,Magdy:2015eda,Redlich:2016vvb,Xu:2016skm,Samanta:2017ohm}. The general question treated in these cited works is to estimate the relevance of the fluctuations due to finite-size effects in the thermodynamic properties of the system. In particular, within the approach of a scalar version of Yukawa model, it is argued in Refs.~\cite{Abreu4,Abreu5} that the reduction of the size of the reservoir which encloses the boson gas might favor the symmetrical phase.

Thus, taking as motivation the discussion done above, in this work we perform an investigation about the influence of the boundaries on the thermodynamic behavior of  Walecka's mean-field theory without quantum correction. In the present study we make use of mean-field approximation for the real scalar and vector fields, which 
could be associated with a first-order estimate of the thermodynamic properties of  hadron matter~\cite{JTheis:1983PRD}. This engenders the interpretation of a thermal gas 
of fermions confined in a reservoir, and interacting with medium constituted of other hadrons. We treat jointly spatial
compactification and the introduction of finite temperature, using generalized Matsubara prescription \cite{Matsubara} and $zeta$-function regularization method  \cite{Abreu1,Abreu2,Abreu3,Luciano1,Khanna,PR2014,EE}. The thermodynamic potential and gap equations can be determined analytically, and the phase structure is analyzed under the change of temperature as well as the number and length of compactified spatial dimensions.

The paper is organized as follows. In Section \ref{Formalism}, we present the model and calculate the relevant thermodynamic quantities using the $zeta$-function regularization approach. In particular, in subsections \ref{SecA} and  \ref{SecB} we introduce the situation without and with boundaries, respectively.  Section \ref{Phase} is devoted to the discussion of the thermodynamics of this system.   Finally, Section \ref{Concluding} presents some concluding remarks.

%%%%%%%%%%%%%%%%%%%%%%%%%%%%%%%%%%%%%%%%%%%%%%%%%%%%%%%%%%%%%%%%%%%%%%%%%%%%%%%%%%%%%%%%
%%%%%%%%%%%%%%%%%%%%%%%%%%%%%%%%%%%%%%%%%%%%%%%%%%%%%%%%%%%%%%%%%%%%%%%%%%%%%%%%%%%%%%%%
\section{The model}
\label{Formalism}
%%%%%%%%%%%%%%%%%%%%%%%%%%%%%%%%%%%%%%%%%%%%%%%%%%%%%%%%%%%%%%%%%%%%%%%%%%%%%%%%%%%%%%%%
%%%%%%%%%%%%%%%%%%%%%%%%%%%%%%%%%%%%%%%%%%%%%%%%%%%%%%%%%%%%%%%%%%%%%%%%%%%%%%%%%%%%%%%%
Let us introduce the effective Lagrangian density of the Walecka model. It describes a Dirac field interacting with 
scalar and vector fields, denoted respectively as $\psi$, $\sigma$ and $\omega$, and is given by 
\begin{eqnarray}
\mathcal{L}& = & \bar{\psi}(i\gamma^{\mu}\partial_{\mu}-m_{\psi}+g_{\sigma}\sigma-g_{\omega}\gamma^{\mu}\omega_{\mu})\psi \nonumber\\
& & +\frac{1}{2}(\partial_{\mu}\sigma\partial^{\mu}\sigma-m_{\sigma}^2\sigma^2)-\frac{1}{4}W^{\mu\nu}W_{\mu\nu} \nonumber\\
& & +\frac{1}{2}m_{\omega}^2\omega_{\mu}\omega^{\mu}
%-V_{\sigma}
,\label{eqII1}
\end{eqnarray}
where $m_{\psi}$, $m_{\sigma}$ e $m_{\omega}$ are the masses of the Dirac, 
scalar and vector fields, respectively; 
$W_{\mu\nu}=\partial_{\mu}\omega_{\nu}-\partial_{\nu}\omega_{\mu}$ is the $\omega$-field strength tensor; 
%$V_{\sigma}$ is the potential associated with the scalar field, which could be chosen to contain self-interacting cubic and quartic terms, i.e. $V_{\sigma}=\frac{1}{3}bM(g_{\sigma}\sigma)^3+\frac{1}{4}c(g_{\sigma}\sigma)^4$; 
and $g_{\sigma }$  ($g_{\omega }$) is the coupling constant for interaction between   
 and Dirac and scalar (vector) field.

In the Lorentz gauge, $\partial_{\mu}\omega^{\mu}=0$, the equations of motion 
obtained from Eq. (\ref{eqII1}) 
%for  $V_{\sigma} = 0 $ 
are
\begin{eqnarray}
[\gamma^{\mu}(\partial_{\mu}-ig_{\omega} \omega_{\mu}) + m_{\psi}-g_{\sigma}\sigma ]\psi & = & 0,
\\ \label{eqII2}
%\end{eqnarray}%
%\begin{eqnarray}%
(\partial_{\mu}\partial^{\mu}+m_{\sigma}^2)\sigma & = & g_{\sigma}\rho_s,
\\ \label{eqII3}
%\end{eqnarray}%
%\begin{eqnarray}%
(\partial_{\mu}\partial^{\mu}+m_{\omega}^2)\omega_{\nu} & = &  ig_{\omega}j_{\nu},\label{eqII4}
\end{eqnarray}
where  $\rho_s=\bar{\psi}\psi$ is the scalar density and $j^{\mu}=\bar{\psi}\gamma ^{\mu}\psi$ the fermion 4-current. Therefore, we have a system of three coupled equations to solve. 

%The Eq.~(\ref{eqII2}) is a type Dirac equation for the $\psi$ field with the meson fields $\sigma$ and%
%$\omega_{\mu}$ included in a minimal substitution. The Eqs.~(\ref{eqII3}) and (\ref{eqII4}) are %
%field equations with massive quanta and $\psi$ spinors currents as source.%
%With the scalar density $\rho_s=\bar{\psi}\psi$, the baryon (vector) density %
%$\rho=j_0=\bar{\psi}\gamma_0\psi$, and the baryon current $\vec{j}=\bar{\psi}\vec{\gamma}\psi$.%
% %

The lowest-order estimate of the thermodynamic properties of the $\psi$-field interacting with other fields can be performed by considering the mean-field approximation. 
It means that we will neglect the fluctuations of the scalar and vector 
fields. In this sense, the $\sigma$ and $\omega$ fields are replaced by 
their classical counterparts, i.e.
\begin{eqnarray} 
\sigma & = & \langle\sigma\rangle, \nonumber \\
%\label{eq105}
%\end{eqnarray}
%\begin{eqnarray} 
\omega & = & \langle\omega^0\rangle,\label{eq106}
\end{eqnarray}
with $\omega^{\mu}=0$ for $\mu\neq0$. Then, in mean-field approximation the
Lagrangian density in Eq.~(\ref{eqII1}) can be rewritten as, 
\begin{eqnarray}
\bar{\mathcal{L}} & = &\bar{\psi}[i\gamma^{\mu}\partial_{\mu}-(m_{\psi}-g_{\sigma}\langle\sigma\rangle)-g_{\omega}\gamma^{0}\langle \omega_{0}\rangle]\psi-\nonumber\\
&&-\frac{1}{2}m_{\sigma}^2\langle\sigma\rangle^2+\frac{1}{2}m_{\omega}^2\langle\omega^{0}\rangle^2, \label{eqII8}
\end{eqnarray}
and the equations of motion in Eqs.~(\ref{eqII2})-(\ref{eqII4}) become
\begin{eqnarray}
[\gamma^{\mu}\partial_{\mu}+(m_{\psi}-g_{\sigma}\langle\sigma\rangle)]\psi & = & g_{\omega}\gamma_{0}\langle\omega^{0}\rangle\psi,
 \label{eqII5} \\
%\end{eqnarray}%
%\begin{eqnarray}%
\langle\sigma\rangle & = & \left(\frac{g_{\sigma}}{m_{\sigma}^2}\right)\rho_s,
\label{eqII6}\\
%\end{eqnarray}%
%\begin{eqnarray}%
\langle\omega_{0}\rangle & = & \left(\frac{g_{\omega}}{m_{\omega}^2}\right)\rho,\label{eqII7}
\end{eqnarray}
where $\rho = j^{0}=\bar{\psi}\gamma ^{0}\psi$ is the fermion density.

To investigate the thermodynamic properties of the model introduced above within imaginary time formalism~\cite{Kapusta,Matsubara}, we assume that the system is in equilibrium at a temperature $T$ and chemical potential
(density) $\mu$. So, we define the grand partition function in a $D$-dimensional Euclidean spacetime at finite temperature $T$ and $d$ compactified spatial dimensions, 
%\begin{widetext}
\begin{eqnarray}
\mathcal{Z}&=&\int\mathcal{D}\psi^{\dagger}\mathcal{D}\psi \nonumber \\
& & \times exp\left\{-\int_0^{\beta}d\tau\prod_{i=1}^{d}\int_0^{L_i}dx_i\int d^{D-\delta}\vec{z}[\bar{\mathcal{L}}_E+\mu j_0]\right\}, \nonumber \\
\label{eqII9}
\end{eqnarray} 
%\end{widetext}
where $\beta=1/T$; $\delta=d+1\leq D$; $\{L_i\}$ are the compactification lengths of the spatial coordinates; $\bar{\mathcal{L}}_E$ is the Lagrangian density given by 
Eq.~(\ref{eqII8}) in Euclidean space; and $\mu$ is the fermion (barionic) chemical potential. 
%Note that the Eq.~(\ref{eqII9}) instructs us %
%to integrate over these field anticommute with each other. 
Finite-size and temperature effects
are taken into account along the prescription described in~\cite{PR2014}: each spatial coordinate $x_i$ is compactified in a length $L_i$
and, as usual, imaginary time is compactified in the range $[0,\beta]$. 
A vector in the $D$-dimensional spacetime is given by
${\bf u}=(\tau, x_1,x_2,\cdots x_d,\vec{z})$,
 where $\tau$ is the imaginary time, $(x_1,x_2,\cdots x_d )$ correspond to the 
 compactified spatial coordinates and $\vec{z}$ a $(D-\delta)$-dimensional vector. The Fourier dual of ${\bf u}$ is a $D$-dimensional vector in momentum space,  ${\bf q}=(k_{\tau},k_{x_1},...,k_{x_d},\vec{p})$, with $\vec{p}$ being the corresponding momentum to $\vec{z}$. As a consequence, in explicit calculations temperature and finite-size
 effects are implemented through the following modifications
in the Feynman rules,
% \begin{widetext}
\begin{eqnarray}
\int\frac{d^Dq}{(2\pi)^D}f(q) & \rightarrow & \frac{1}{\beta L_i\cdot\ldots\cdot L_d}
\nonumber \\
& & \times \sum_{l,\{n_i\}=-\infty}^{\infty}\int\frac{d^{D-\delta}p}{(2\pi)^{D-\delta}}f(\omega_l,\{ \omega_{n_i} \},p) , \nonumber \\
\label{eqII10}
\end{eqnarray}
% \end{widetext}
where  $\{n_i\} \equiv n_1, \ldots , n_d $; $\{\omega _i\} \equiv \omega_1, \ldots , \omega_d $; in the right-hand side, we have performed the replacements 
\begin{eqnarray}
k_{\tau}\rightarrow\omega_l&=&\frac{2\pi}{\beta}\left(l+\frac{1}{2}\right)-i\mu;\hspace{0.25cm}l=0,\pm1,\pm2,...,\nonumber\\
k_{x_i}\rightarrow\omega_{n_i}&=&\frac{2\pi}{L}(n_{i}+c);\hspace{0.75cm}n_{i}=0,\pm1,\pm2,...,\nonumber
\end{eqnarray}
with $c=0$ or $c=1/2$ for periodic or antiperiodic 
spatial boundary conditions, respectively. In the present study we use the   antiperiodic spatial boundary conditions. 

Then, after the integration of the fields $\psi^{\dagger}$ and $\psi$, we can obtain from Eq.~(\ref{eqII9}) the thermodynamic potential,
%\begin{widetext}
\begin{eqnarray}
U(T,\{L_i\},\mu_{eff})& \equiv & \frac{1}{\beta} \ln{\mathcal{Z}} \nonumber \\
& = & \frac{V}{2}\left[m_{\sigma}^2\langle\sigma\rangle^2-m_{\omega}^2\langle\omega^{0}\rangle^2\right]-
\nonumber \\
& &-\frac{\gamma V}{\beta L_i\cdot\ldots\cdot L_d}Y^{'}(0), 
\label{eqII11}
\end{eqnarray}
%\end{widetext}
where $\gamma$ is the degeneracy factor (we adopt here $\gamma=4$ \cite{JTheis:1983PRD}), $V$ is the volume, and $Y(s)$ is the multiple sum obtained after performing the integration over the  $(D - \delta )$-dimensional momentum vector remaining from the prescription in Eq. (\ref{eqII10}):
%\begin{eqnarray}%
%Y(s)=\sum_{l,n_1,\ldots,n_d=-\infty}^{\infty}\int\frac{d^{D-d}p}{(2\pi)^{D-d}}\left\{\left[\frac{2\pi}{\beta}\left(l+\frac{1}{2}\right)-i\mu_{eff}\right]^2+\left[\frac{2\pi}{L_1}\left(n_1+\frac{1}{2}\right)\right]^2+...+\left[\frac{2\pi}{L_d}\left(n_d+\frac{1}{2}\right)\right]^2+p^2+m^2_{eff}\right\}^{-s}\label{eqII12}%
%\end{eqnarray}%
%\end{widetext}%
%where $Y^{'}(s)$ stands for the derivative of $Y(s)$ with respect to the argument $s$,%
%$\mu_{eff}$ and $m_{eff}$ are respectively, the effective chemical potential and mass %
%of the baryons field $\psi$; they read,%
%\begin{subeqnarray}\label{eq:sub}%
%\slabel{sub1}m_{eff}&=&m_{\psi}-g_{\sigma}\langle\sigma\rangle,\label{eqII13}\\%
%\slabel{sub2}\mu_{eff}&=&\mu-g_{\omega}\langle\omega^0\rangle,\label{eqII14}%
%\end{subeqnarray} %
%The modifications of the scalar and vector mean fields modify the masses and chemical %
%potential of the baryons (Nucleons for example) in the hot and dense hadronic medium. Therefore, %
%in the lowest order the interactions are manifested in the Eqs.~(\ref{eqII13}a) and (\ref{eqII14}b).%
%The integral in the above equation is performed using dimensional regularization; we get,%
%\begin{widetext}
\begin{eqnarray}
Y(s)& = & J(s,\delta)
\sum_{l,n_1,\ldots,n_d=-\infty}^{\infty}\left\{\left[\frac{2\pi}{\beta}\left(l+\frac{1}{2}\right)-i\mu_{eff}\right]^2 \right.
\nonumber \\
& & + \left. \sum _{i=1} ^{d} \left[\frac{2\pi}{L_i}\left(n_i+\frac{1}{2}\right)\right]^2 + m^2_{eff}\right\}^{-s+\frac{D-\delta}{2}},
\label{eqII15}
\end{eqnarray}
%\end{widetext}
with 
\begin{eqnarray}
J(s,\delta)=\frac{1}{(4\pi)^{(D-\delta)/2}}\frac{\Gamma\left(s-\frac{D-\delta}{2}\right)}{\Gamma(s)}. \label{eqII16}
\end{eqnarray}
$Y^{'}(s)$ in Eq. (\ref{eqII15}) represents the derivative of $Y(s)$ with respect to the argument $s$;  $m_{eff}$ and $\mu_{eff}$ are respectively the effective mass and chemical potential of the fermion field, and are given by
\begin{eqnarray}%\label{eq:sub}
m_{eff}&=&m_{\psi}-g_{\sigma}\langle\sigma\rangle,\label{eqII13}\\
\mu_{eff}&=&\mu-g_{\omega}\langle\omega^0\rangle.\label{eqII14}
\end{eqnarray} 
Then,  modifications of the scalar and vector mean fields [whose allowed values are obtained from the solutions of gap equations in Eqs. (\ref{eqII6}) and (\ref{eqII7})] change the mass and chemical potential of the fermions. We can interpret this as follows: the mass and density of fermions (for instance nucleons) are effectively modified due to their interactions with the hot and dense medium in which they are immersed.

It is relevant to notice that the multiple sum in Eq. (\ref{eqII15}) is the well known Epstein-Hurwitz inhomogeneous $zeta$-function \cite{EE}, whose analytical continuation valid in the whole complex $\nu$-plane has following representation, 
\begin{widetext}
\begin{eqnarray}
A^{C^2}_\delta(\nu,\{a_i\},\{b_i\})&=&
\sum_{\{n_i\}=-\infty}^{\infty}\left[\sum_{i=1}^{\delta} a_i(n_i-b_i)^2 + C^2 \right]^{-\nu} \nonumber \\
& = & \frac{\pi^{\delta/2}}{\sqrt{a_1\cdot\ldots\cdot a_\delta}\Gamma(\nu)}\left\{ 
\Gamma\left(\nu-\frac{\delta}{2}\right)C^{\delta-2\nu} +2\sum_{i=1}^{\delta}\sum_{n_i=1}^{\infty}cos(2\pi n_i b_i)\left(\frac{\pi n_i}{\sqrt{a_i} C}\right)^{\nu-\frac{\delta}{2}} K_{\nu-\frac{\delta}{2}}\left(\frac{2\pi n_i C}{\sqrt{a_i}}\right) 
\right. \nonumber\\
& & + 2^2\sum_{i<j=1}^{\delta}\sum_{n_i,n_j=1}^{\infty}cos(2\pi n_i b_i)cos(2\pi n_j b_j)\left(\frac{\pi }{C^2}\sqrt{\frac{n_i^2}{a_i}+\frac{n_j^2}{a_j}}\right)^{\nu-\frac{\delta}{2}} K_{\nu-\frac{\delta}{2}}\left(2\pi C \sqrt{\frac{n_i^2}{a_i} +\frac{n_j^2}{a_j}}\right) \nonumber\\
& & \left. + \cdots + 2^\delta \sum_{n_1,\ldots,n_\delta=1}^{\infty} \prod_{i=1}^{\delta} \left[cos(2\pi n_i b_i) \right] \left(\frac{\pi }{C^2}\sqrt{\sum_{i=1}^{\delta }\frac{n_i^2}{a_i}}\right)^{\nu-\frac{\delta}{2}} K_{\nu-\frac{\delta}{2}}\left(2\pi C \sqrt{\sum_{i=1}^{\delta } \frac{n_i^2}{a_i}} \right) \right\},
\label{eqII18}
\end{eqnarray} 
\end{widetext}
where $K_{\nu}$ is the modified Bessel function of the second kind.

Thus, the grand thermodynamic potential can be obtained by taking the derivative of the  function $Y(s)$ in Eq.~(\ref{eqII18}), and replacing the result into Eq.~(\ref{eqII11}). In Eq.~(\ref{eqII11}) we have $\nu=s-(D-\delta)/2$, which engenders the label of the Bessel functions equal to $\nu-\delta/2 = s-D/2$. Therefore, for $D=4$ the Bessel functions are  $K_{s-2}$ independently of the value of $\delta$.

In addition, to study the thermodynamic behavior of the system we need to 
consider the state equations of mean fields $\langle\sigma\rangle$ and $\langle\omega^0\rangle$,  
\begin{eqnarray}
\frac{\partial U}{\partial\langle\sigma\rangle}&=&0,\label{eqII19} \\
\frac{\partial U}{\partial\langle\omega^{0}\rangle}&=&0.\label{eqII20}
\end{eqnarray}  
The solutions of these equations give the values of $\langle\sigma\rangle$ and 
$\langle\omega^0\rangle$ corresponding to the
extrema of the grand thermodynamic potential $U$ where the system reaches the 
the equilibrium configuration. In this sense, we are interested in the fermion effective mass $m_{eff}=m_{eff}(T,\mu_{eff},\{L_i\})$ defined in Eq.~(\ref{eqII13}), which will be a  $(T,\mu_{eff},\{L_i\})$-dependent order
parameter which governs the phase diagram of the model. 

%%%%%%%%%%%%%%%%%%%%%%%%%%%%%%%%%%%%%%%%%%%%%%%%%%%%%%%%%%%%%%%%%%%%%%%%%%%%%%%%%%%%%%%%
%%%%%%%%%%%%%%%%%%%%%%%%%%%%%%%%%%%%%%%%%%%%%%%%%%%%%%%%%%%%%%%%%%%%%%%%%%%%%%%%%%%%%%%%
\subsection{System without spatial boundaries ($d=0$)}\label{SecA}
%%%%%%%%%%%%%%%%%%%%%%%%%%%%%%%%%%%%%%%%%%%%%%%%%%%%%%%%%%%%%%%%%%%%%%%%%%%%%%%%%%%%%%%%
%%%%%%%%%%%%%%%%%%%%%%%%%%%%%%%%%%%%%%%%%%%%%%%%%%%%%%%%%%%%%%%%%%%%%%%%%%%%%%%%%%%%%%%%

For completeness, we begin with the usual case where the system is in the absence of boundaries. It means that we use the recurrence formula in Eq. (\ref{eqII18}) with $D=4$ and $\delta=1$, identifying $a_1=\left(\frac{2\pi}{\beta}\right)$ 
and $b_1=-i\frac{\mu_{eff}\beta}{2\pi}+\frac{1}{2}$, and after that we replace the expression obtained for $Y(s)$ in Eq. (\ref{eqII15}). 
Nevertheless, we must perform the pole structure analysis in the calculations of the derivative of $Y(s)$ with respect to $s$, for $s\rightarrow0$,  before use it in the expression of thermodynamic potential (\ref{eqII11}). Accordingly, it can be remarked that for any regular function $G(s)$, have, 
$\lim_{s\rightarrow0}(d/ds)[G(s)/\Gamma(s)]=G(0)$.

Then, after performing the derivative of $Y(s)$ given in (\ref{eqII15}) with respect to $s$ for $\epsilon\rightarrow 0$, but with the observations mentioned above in mind, we can rewrite the thermodynamic potential as, 
\begin{eqnarray}
\frac{U(T,\mu_{eff})}{V}&=& \frac{1}{2}m_{\sigma}^2\langle\sigma\rangle^2 -\frac{1}{2}m_{\omega}^2\langle\omega^{0}\rangle^2 - U_{vac} +
\nonumber \\
& & \frac{\gamma}{\pi^2}\sum^{\infty}_{l=1}(-1)^{l}\cosh(\beta l\mu_{eff})\left(\frac{m_{eff}}{l\beta}\right)^{2} \nonumber\\
&&\times K_{2}(l\beta m_{eff}) ,
 \label{eqIIA1}
\end{eqnarray}
where $U_{vac}$ is the vacuum fluctuation energy per unit volume, i.e. the quantum correction, associated to the first term in the second line of Eq.~(\ref{eqII18}). This contribution can be better understood from the discussions available in Refs.~\cite{Chin,Freedman1,MATSUI}: since the scalar interaction effectively changes the fermion mass from $m_{\psi}$ to $m_{eff} = m_{\psi} - g_{\sigma} \langle\sigma\rangle$, it induces an energy density shift of the vacuum of magnitude 
\begin{eqnarray}
U_{vac} & \propto &  \left[ \frac{1}{(4\pi)^{\frac{D}{2}-1}}\Gamma \left( - \frac{D}{2} \right) m_{eff}^D \right]_{D=4} 
\nonumber \\
& & -  \left[ \frac{1 }{(4\pi)^{\frac{D}{2}-1}}\Gamma \left( - \frac{D}{2} \right) m_{\psi}^D \right]_{D=4},
 \label{eqIIA2}
\end{eqnarray}
where the second term in right hand side of equation above has been introduced in order to eliminate the physically meaningless constant shift of the energy.
Clearly this expression contains divergences, which can be  canceled by renormalization procedure as done in Refs.~\cite{Chin,MATSUI}. However, it is relevant to notice that  in the present work we are interested in the finite-size effects on the phase structure of the model introduced in Ref.~\cite{JTheis:1983PRD}. In other words, we investigate the influence of boundaries on the thermodynamic behavior of the Walecka's mean-field theory, without quantum correction. 
Thus, henceforth we will omit  in the calculations  the  term in Eq.~(\ref{eqIIA2}).  In this sense, the present analysis  can be considered as a starting point for further work in which the quantum corrections might be considered.

Thus, the relevant thermodynamic properties can be derived from the grand thermodynamic potential in Eq.~(\ref{eqIIA1}). The substitution of Eq.~(\ref{eqIIA1}) in Eqs.~(\ref{eqII19}) and (\ref{eqII20}) allows us to rewrite the gap equations as 
\begin{eqnarray}
\langle\sigma\rangle&=&\frac{g_{\sigma}}{m_{\sigma}^2}\rho_s,\label{eqIIA3}\\
\langle\omega^0\rangle&=&-\frac{{\omega}}{m_{\omega}^2}\rho,\label{eqIIA4}
\end{eqnarray}
where $\rho_s$ and $\rho$ are the scalar and number densities, respectively, given by, 
\begin{eqnarray}
\rho_s&=&\frac{m_{eff}^2}{\pi^2\beta}\sum^{\infty}_{l=1}a_l\cosh(l\beta\mu_{eff})K_{1}(l\beta m_{eff}),\label{eqIIA5}\\
\rho&=&\frac{m_{eff}^2}{\pi^2\beta}\sum^{\infty}_{l=1}b_l\sinh(l\beta\mu_{eff})K_{2}(l\beta m_{eff}),\label{eqIIA6}
\end{eqnarray}
with $a_l=\frac{(-1)^{l-1}}{l}$ and $b_l=-2\frac{(-1)^{l-1}}{l}$. 

\bigskip

%%%%%%%%%%%%%%%%%%%%%%%%%%%%%%%%%%%%%%%%%%%%%%%%%%%%%%%%%%%%%%%%%%%%%%%%%%%%%%%%%%%%%%%%
%%%%%%%%%%%%%%%%%%%%%%%%%%%%%%%%%%%%%%%%%%%%%%%%%%%%%%%%%%%%%%%%%%%%%%%%%%%%%%%%%%%%%%%%
\subsection{System with compactified spatial dimensions ($d=1,2,3$)}\label{SecB}
%%%%%%%%%%%%%%%%%%%%%%%%%%%%%%%%%%%%%%%%%%%%%%%%%%%%%%%%%%%%%%%%%%%%%%%%%%%%%%%%%%%%%%%%
%%%%%%%%%%%%%%%%%%%%%%%%%%%%%%%%%%%%%%%%%%%%%%%%%%%%%%%%%%%%%%%%%%%%%%%%%%%%%%%%%%%%%%%%
Now we present the model under question with the presence of boundaries, in the  scenarios of $d=1,2$ and 3 compactified spatial coordinates.

We start by considering the simplest case, which is just one spatial compactfication, $d=1$, corresponding to a reservoir in the form of an infinite hollow slab  of thickness $L_1 \equiv L$, in with the system at equilibrium 
and at a temperature $\beta^{-1}$.  Then, to obtain the thermodynamic potential we use Eqs.~(\ref{eqII15}) and (\ref{eqII18}), 
with $D=4$ and $\delta=2$, and introduce the notation $a_1=\left(\frac{2\pi}{\beta}\right)$, 
$a_2=\left(\frac{2\pi}{L}\right)$, $b_1=-i\frac{\mu_{eff}\beta}{2\pi}+\frac{1}{2}$ and $b_2=\frac{1}{2}$. 
Then, proceeding similarly as the way to find Eqs.~(\ref{eqIIA1}), (\ref{eqIIA3}) and (\ref{eqIIA4}), the grand thermodynamic potential in Eq.~(\ref{eqIIA1}) becomes,
\begin{widetext}
\begin{eqnarray}
\frac{U(T,\mu_{eff},L)}{V}&=& \frac{1}{2}m_{\sigma}^2\langle\sigma\rangle^2 -\frac{1}{2}m_{\omega}^2\langle\omega^{0}\rangle^2 +U_{vac} \nonumber\\
& & + \sum_{l=1}^{\infty}\hspace{0.03cm}(-1)^{l}\hspace{0.03cm}cosh(\hspace{0.03cm}l\hspace{0.03cm}\beta\hspace{0.03cm}\mu_{eff}\hspace{0.03cm})\hspace{0.03cm}\left(\frac{m_{eff}}{l\hspace{0.03cm}\beta}\right)^{2}\hspace{0.03cm}K_{2}\left(\hspace{0.03cm}l\hspace{0.03cm}m_{eff}\hspace{0.03cm}\beta\hspace{0.03cm}\right)\hspace{0.075cm}+\hspace{0.075cm}\sum_{n=1}^{\infty}\hspace{0.03cm}(-1)^{n}\hspace{0.03cm}\left(\frac{m_{eff}}{n\hspace{0.01cm}L}\right)^{2}\hspace{0.03cm}K_{2}\left(\hspace{0.03cm}n\hspace{0.03cm}m_{eff}\hspace{0.03cm}L\hspace{0.03cm}\right)\hspace{0.05cm}+\nonumber\\
&& +\sum_{l,n=1}^{\infty}(-1)^{l+n-1}cosh(l\beta\mu_{eff})\left(\frac{m_{eff}}{\sqrt{\beta^2l^2+L^2n^2}}\right)^{2}K_{2}\left(m_{eff}\sqrt{\beta^2l^2+L^2n^2}\right). 
\label{eqIIB1}
\end{eqnarray} 
\end{widetext} 

Then, the use of Eq.~(\ref{eqIIB1}) in the gap equations (\ref{eqII19}) and (\ref{eqII20}) and after some mathematical manipulations
yield similar gap equations shown in Eqs.~(\ref{eqIIA3}) and~(\ref{eqIIA4}), but with the scalar and number densities in the present case with $d=1$ being respectively given by 
%\begin{eqnarray}%
%\langle\sigma\rangle&=&\frac{g_{\sigma}}{m_{\sigma}^2}\rho_s^{L},\label{eqIIB2}\\%
%\langle\omega^0\rangle&=&-\frac{g_{\omega}}{m_{\omega}^2}\rho^{L},\label{eqIIB3}%
%\end{eqnarray}%
%where $\rho_s^{L}$ and $\rho^{L}$ are respectively the number and scalar densities in the case with $d=1$, %
\begin{widetext}
\begin{eqnarray}
\rho_s &=&A\left[ \sum_{i=0}^{1}\sum_{n_i=1}^{\infty}\frac{d_{n_i}}{L_i}cosh(\delta_{0i}n_iL_i\mu_{eff})K_{1}(n_im_{eff}L_i)+2\sum_{n,l=1}^{\infty}c_{n,l}cosh(l\beta\mu_{eff})K_{1}\left(m_{eff}\sqrt{\sum_{i=0}^1L^2_in^2_i}\right)\right],\label{eqIIB4}\\
\rho &=&-4\pi^2A^2\left[\sum_{l=1}^{\infty}\frac{b_l}{2\beta}senh(l\beta\mu_{eff})K_{2}(lm_{eff}\beta)-2\beta\sum_{l,n=1}^{\infty}b_{l,n}senh(n\beta\mu_{eff})K_{2}\left(m_{eff}\sqrt{\sum_{i=0}^1L^2_in^2_i}\right)\right],\label{eqIIB5}
\end{eqnarray} 
\end{widetext}
where $A=\frac{m_{eff}}{\pi^2}$, $d_{n_0}=a_l$, $d_{n_1}=c_n$, $c_{n,l}=\frac{(-1)^{n+l}}{\sqrt{\sum_{i=0}^1L^2_in^2_i}}$, $b_{n,l}=\frac{2(-1)^{n+l}}{\sqrt{\sum_{i=0}^1L^2_in^2_i}}$ ; also, we use the notation $L_0 \equiv \beta$, $L_1 \equiv L$,  and $n_0 \equiv l$ and $n_1 \equiv n$.
%%%%%%%%%%%%%%%%%%%%%%%%%%%%%%%%%%%%%%%%%%%%%%%%%%%%%%%%%%%%%%%%%%%%%%%%%%%%%%%%%%%%%%%%
%%%%%%%%%%%%%%%%%%%%%%%%%%%%%%%%%%%%%%%%%%%%%%%%%%%%%%%%%%%%%%%%%%%%%%%%%%%%%%%%%%%%%%%%

We also analyze the present approach with two compactfied spatial coordinates ($d=2$), in which the reservoir has the geometry of 
an hollow infinitely long wire with a retangular cross-section. 
In this context we return to Eqs.~(\ref{eqII15}) and (\ref{eqII18}) with $D=4$ and $\delta = 3$, and introduce the quantities  
$a_1=\left(\frac{2\pi}{\beta}\right)$, $a_2=\left(\frac{2\pi}{L_1}\right)$, $a_3=\left(\frac{2\pi}{L_2}\right)$
$b_1=-i\frac{\mu_{eff}\beta}{2\pi}+\frac{1}{2}$, $b_2=\frac{1}{2}$ and $b_3=\frac{1}{2}$. 
Then, proceeding similarly as the way to find Eqs.~(\ref{eqIIA1}), (\ref{eqIIA3}) and (\ref{eqIIA4}), the  $(L_1,L_2)$-dependent expressions for grand thermodynamic potential in Eq.~(\ref{eqIIA1}), gap equations in Eqs.~(\ref{eqII19}) and (\ref{eqII20}) and number and scalar densities can be generated after some manipulations. For brevity, we omit these expressions as the reader can obtain them for himself.

%%%%%%%%%%%%%%%%%%%%%%%%%%%%%%%%%%%%%%%%%%%%%%%%%%%%%%%%%%%%%%%%%%%%%%%%%%%%%%%%%%%%%%%%
%%%%%%%%%%%%%%%%%%%%%%%%%%%%%%%%%%%%%%%%%%%%%%%%%%%%%%%%%%%%%%%%%%%%%%%%%%%%%%%%%%%%%%%%

Finally, looking at the case of three compactfied spatial coordinates  ($d=3$), the reservoir has the form of parallelepipedal box of volume $L_1\times L_2\times L_3$. Once more, $Y(s)$ can be be obtained by taking $D=4$ and $\delta = 4$
in Eqs.~(\ref{eqII15}) and (\ref{eqII18}),
and introducing the quantities $a_1=\left(\frac{2\pi}{\beta}\right)$, $a_2=\left(\frac{2\pi}{L_1}\right)$, 
$a_3=\left(\frac{2\pi}{L_2}\right)$, $a_3=\left(\frac{2\pi}{L_3}\right)$, 
$b_1=-i\frac{\mu_{eff}\beta}{2\pi}+\frac{1}{2}$, $b_2=\frac{1}{2}$, $b_3=\frac{1}{2}$, 
and $b_4=\frac{1}{2}$. 
Hence, $(L_1,L_2,L_3)$-dependent expressions for the thermodynamic potential, scalar and number densities, pressure, entropy and other thermodynamic quantities are determined by similar procedures to those described above.

In next Section we will carry out the discussion of the thermodynamics of the model in these different situations. 

%%%%%%%%%%%%%%%%%%%%%%%%%%%%%%%%%%%%%%%%%%%%%%%%%%%%%%%%%%%%%%%%%%%%%%%%%%%%%%%%%%%%%%%%
%%%%%%%%%%%%%%%%%%%%%%%%%%%%%%%%%%%%%%%%%%%%%%%%%%%%%%%%%%%%%%%%%%%%%%%%%%%%%%%%%%%%%%%%
\section{Phase Structure and Comments}\label{Phase}
%%%%%%%%%%%%%%%%%%%%%%%%%%%%%%%%%%%%%%%%%%%%%%%%%%%%%%%%%%%%%%%%%%%%%%%%%%%%%%%%%%%%%%%%
%%%%%%%%%%%%%%%%%%%%%%%%%%%%%%%%%%%%%%%%%%%%%%%%%%%%%%%%%%%%%%%%%%%%%%%%%%%%%%%%%%%%%%%%

Now we are able to analyze the thermodynamic behavior of the system previously introduced.

In the present approach, we consider the case in which the system is at effective chemical equilibrium, i.e. $\mu_{eff}=0$. It can be seen from Eqs. (\ref{eqIIA5}) and (\ref{eqIIA6}) that this constraint engenders  $\rho=0$, and therefore yields vanishing solution for the $\langle\omega^{0}\rangle$ field in all cases of compactified spatial dimensions. It means that in the present analysis the numbers of fermion particles and antiparticles are equal.  
Therefore, we focus on the thermodynamics of the system as a function of the solutions of the gap equation  (\ref{eqIIA5}) for $ \langle \sigma \rangle = \langle \sigma \rangle (T, L_i)$ or $m_{eff} =m_{eff} (T, L_i)$ at $\langle\omega^0 \rangle = 0 $.

%The results below represent some consequences from the gap equations %
%(\ref{eqIIA4}a) and (\ref{eqIIB4}a) and analogous cases in $d=2,3$ %
%under changes of the relevant parameters of the model (For $L_i=L$ as $i=1,2$ or $3$). %

For convenience,  all physical quantities are scaled by units of mass 
of the fermion field $\psi$, $m_{\psi}$:  
\begin{eqnarray}
\begin{array}{lccc}
\frac{U}{m_{\psi}^4} &\rightarrow & U, \;\; \frac{T}{m_{\psi}}\rightarrow T,\;\; \frac{\mu}{m_{\psi}}
\rightarrow\mu, \;\; \frac{\sigma}{m_{\psi}}\rightarrow\sigma,\\
\frac{m_{\sigma}}{m_{\psi}} & \rightarrow & m_{\sigma},\;\; \frac{m_{eff}}{m_{\psi}}\rightarrow m_{eff},\;\; Lm_{\psi}\rightarrow L .
\end{array}\label{eqIII1}
\end{eqnarray}

The discussion about the phase structure of the present model is accomplished under changes of the relevant parameters, paying attention to the influence of the number of compactified spatial dimensions. In this sense, the phase transitions are analyzed and compared with the system in the situations of one, two and three compactified spatial dimensions. To make this comparison more accessible, the length of all spatial compactified coordinates is set to be the same, i.e. $L_i \equiv L$.  This choice reduces the system to the following scenarios for $d=1,2$ and 3: confined between two parallel planes a distance $L$ apart; confined to an infinitely long cylinder having a square transversal section of area $L^2$; and to a cubic box of volume $L^3$.

%%%%%%%%%%%%%%%%%%%%%%%%%%%%%%%%%%%%%%%%%%%%%%%%%%%%%%%%%%%%%%%%%%%%%%%%%%%%%%%%%%%%%%%%
%%%%%%%%%%%%%%%%%%%%%%%%%%%%%%%%%%%%%%%%%%%%%%%%%%%%%%%%%%%%%%%%%%%%%%%%%%%%%%%%%%%%%%%%
\subsection{System without spatial boundaries ($d=0$)}\label{ResultsSecA}

For completeness, we begin by a qualitative analysis of the behavior of the effective mass $m_{eff}$ under changes of the relevant parameters but without the presence of boundaries, similarly to the scenario described in Ref.~\cite{JTheis:1983PRD}\footnote{We notice that we use a different scaling and notation with respect to Ref.~\cite{JTheis:1983PRD}.}. In what follows, including the figures, all parameters are understood to be redefined by the scaling in Eq.~(\ref{eqIII1}).

%\begin{figure}[th]%
%\centering%
%\includegraphics[{height=6.5cm,width=6.5cm}]{MassaEfetiva1.eps}%
%\caption{Plot of effective mass in Eq.~(\ref{eqII13}a) as a function of temperature, at chemical %
%equilibrium. We fix $m_{\sigma}=0.53$, with full and dashed lines  representing respectively %
%$g_{\sigma}=16.00$ and $g_{\sigma}=26.00$. }%
%\label{fig:MassaEfetiva1}%
%\end{figure}%
%%
%We can to see in Fig.~\ref{fig:MassaEfetiva1} a sudden drop in %
%$m_{eff}$ at temperature $T\approx0.15$ for $g_{\sigma}=26.00$\footnote{The choice of this value %
%of the coupling constant has a pedagogical role to present the results.} and $T\approx0.20$ for %
%$g_{\sigma}=16.00$, where there is a transition from $m_{eff}=0.98$ to $m_{eff}=0.20$ what ocurrs %
%in interval of $\Delta T\approx0.045-0.075$ only. 

\begin{figure}[th]
\centering
\includegraphics[{width=8.0cm}]{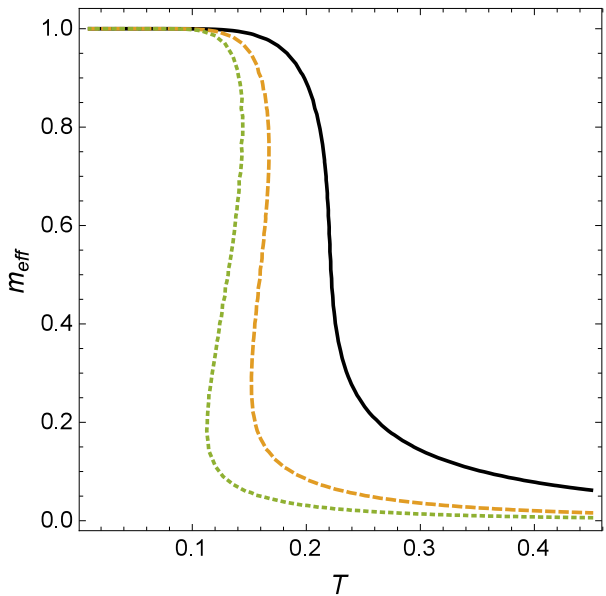}
\caption{Plot of values of $m_{eff}$ that are solutions of the gap equation in Eq. (\ref{eqIIA3}) for $ d=0$ as a function of temperature, at chemical 
equilibrium. We fix $m_{\sigma}=0.53$. Full, dashed and dotted lines represent respectively the cases for $g_{\sigma}=8.00$, $g_{\sigma}=16.00$ and $g_{\sigma}=26.00$.}
\label{fig:MassaEfetiva1}
\end{figure}

\begin{figure}[th]
\centering
\includegraphics[{width=8.0cm}]{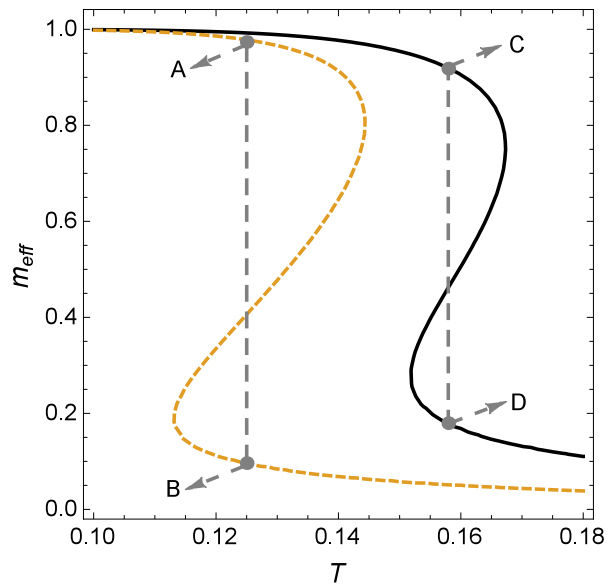}
\caption{Same as in Fig.~\ref{fig:MassaEfetiva2}, but with full and dashed lines representing respectively the cases for $g_{\sigma}=16.00$ and $g_{\sigma}=26.00$,  and with the temperature axis in the range near the transition point.}
\label{fig:MassaEfetiva2}
\end{figure}

In Figs.~\ref{fig:MassaEfetiva1} and~\ref{fig:MassaEfetiva2} are plotted the values of $m_{eff}$ that are solutions of the gap equation in Eq. (\ref{eqIIA3}) as function of temperature, by taking arbitrary values of the coupling constant $g_{\sigma}$. 
As discussed in Ref.~\cite{JTheis:1983PRD}, the increasing of the magnitude of the interaction between the $\sigma$ and $\psi$ fields  makes the system
to undergo a phase transition at a smaller critical temperature. This can be understood from the role of the $\sigma$ field in binding
the fermion particles: the increasing of the magnitude of
attractive interaction between the fermions makes them more strongly bound, reducing the effective mass. Furthermore,   
the nature of the phase transition strongly depends on the strength of the coupling constant $ g_{\sigma} $. For the case with lowest $ g_{\sigma} $ the dependence of $m_{eff}$ on $T$ is continuous; while for larger $ g_{\sigma} $ the curves cross twice, characterizing a discontinuous phase transition. 

Specifically, in Fig.~\ref{fig:MassaEfetiva2}  is plotted the solutions for  $m_{eff}$ as a function of temperature for larger values of $g_{\sigma}$ which engender a first order transition, with the temperature axis in the range near the transition point. The points $\overline{AB}$ and $\overline{CD}$ 
indicate the ordered pairs of coordinates $(m_{eff}, T)$ of the intersections that give the values of the critical temperature and $m_{eff} $ at both end points of the mixed phase. In the case of $g_{\sigma}=26.00$
the jump occurs at $T \approx 0.125-0.126$, where the effective mass changes from $\approx0.98$ to $\approx0.1$; 
while for $g_{\sigma}=16.00$  the phase transition is at $T\approx 0.158-0.159$, where $m_{eff}$ goes from $\approx0.90$ to $\approx0.20$.
At high temperatures, the system behaves like an almost-free zero-mass fermion gas.

It is worthy mentioning that a more detailed study of the scenario above for several values of $g_{\sigma}$ gives the following result: for $g_{\sigma} < 9.8 $ the effective mass is  smooth in the temperature, whereas for $g_{\sigma} > 9.8 $ a phase transition of first order takes place.

\begin{figure}[th]
\centering
\includegraphics[{width=8.0cm}]{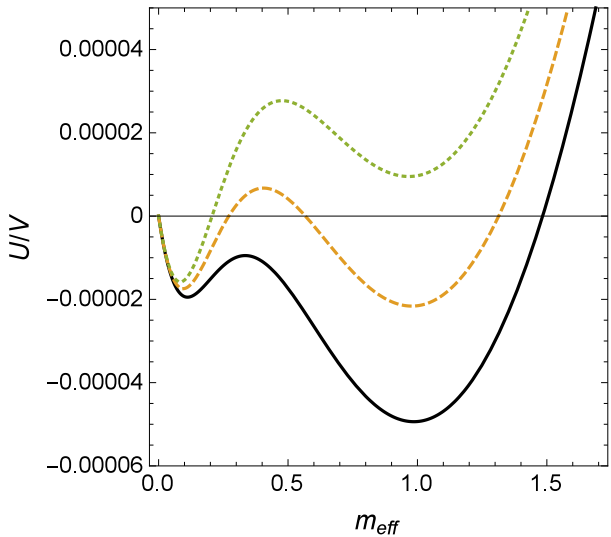}
\caption{Plot of normalized thermodynamic potential density in Eq.~(\ref{norm_pot}) for $d=0$ as a function of effective mass and 
at chemical equilibrium. We fix $m_{\sigma}=0.53$ and $g_{\sigma}=26.00$. Full, dashed and dotted 
lines represent the cases for $T=0.120$, $T=0.125$ and $T=0.130$, respectively.}
\label{fig:PotencialBulk1}
\end{figure}

\begin{figure}[th]
\centering
\includegraphics[{width=8.0cm}]{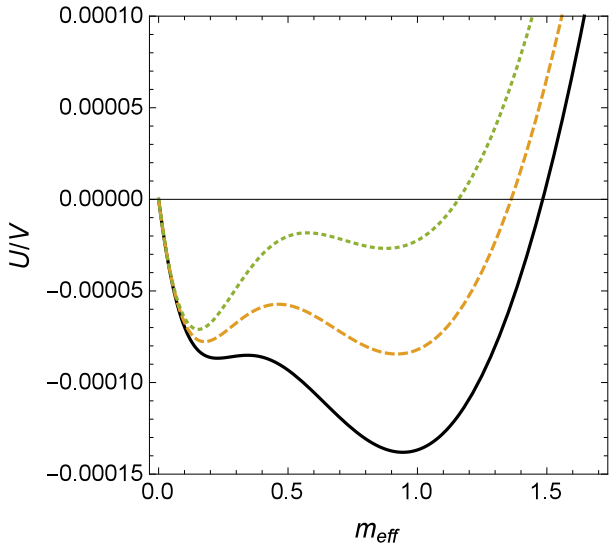}
\caption{Same as in Fig.~\ref{fig:PotencialBulk1}, but with $g_{\sigma}=16.00$. Full, dashed and dotted 
lines represent the cases for $T=0.153$, $T=0.158$ and $T=0.163$, respectively.}
\label{fig:PotencialBulk2}
\end{figure}

To complete the characterization of the thermodynamics of this
system we examine the thermodynamic potential density, which will be normalized with respect to reference state $U(m_{eff}=0)$:
\begin{eqnarray}
\frac{1}{V} \left[ U(m_{eff})-U(m_{eff}=0) \right] \rightarrow \frac{U(m_{eff})}{V}.
\label{norm_pot}
\end{eqnarray}
Thus, in Figs.~(\ref{fig:PotencialBulk1}) 
and~(\ref{fig:PotencialBulk2}) the normalized thermodynamic potential density $U(T,\mu_{eff}=0)/V$, obtained by using Eq.~(\ref{eqIIA1}) in (\ref{norm_pot}), is plotted as a function of effective mass for $g_{\sigma}=26.00$ and $g_{\sigma}=16.00$, 
respectively, and for $m_{\sigma}=0.53$. The discontinuous phase transition is clearly seen as $T$ increases; the absolute minimum of the potential 
is displaced to a smaller value of $m_{eff}$, as $T$ increases. As suggested in Fig.~\ref{fig:MassaEfetiva2}, this first-order phase transition occurs at smaller  critical temperatures as the magnitude of interaction between the $\sigma$ and $\psi$ fields increases. For larger values of temperature, the effective mass a decreases smoothly . 

%The fixed coupling constants $g_{\sigma}$ and the mass sigma field $m_{\sigma}$, %
%$\left(\frac{g_{\sigma}}{m_{\sigma}}\right)^2=365$, to give %
%$\epsilon_{bind}=-15.75$ MeV and $k_F=1.42$ fm$^{-1}$, also, to found the result %
%the of equilibrium density $\rho_{eq}=0.19$ fm$^{-3}$ and the nuclear incompressibility %
%$\kappa\approx550$ MeV, thus, unrealistically large \cite{}. For this case the phase %
%trasition ocurrs in  $T\approx0.2$ when the effective mass from $m_{eff}=0.75$ %
%to $m_{eff}=0.35$ in according to the literature \cite{JTheis:1983PRD}.%

%%%%%%%%%%%%%%%%%%%%%%%%%%%%%%%%%%%%%%%%%%%%%%%%%%%%%%%%%%%%%%%%%%%%%%%%%%%%%%%%%%%%%%%%
%%%%%%%%%%%%%%%%%%%%%%%%%%%%%%%%%%%%%%%%%%%%%%%%%%%%%%%%%%%%%%%%%%%%%%%%%%%%%%%%%%%%%%%%

\subsection{System with compactified spatial dimensions ($d=1,2,3$)}\label{ResultsSecB}
%%%%%%%%%%%%%%%%%%%%%%%%%%%%%%%%%%%%%%%%%%%%%%%%%%%%%%%%%%%%%%%%%%%%%%%%%%%%%%%%%%%%%%%%
%%%%%%%%%%%%%%%%%%%%%%%%%%%%%%%%%%%%%%%%%%%%%%%%%%%%%%%%%%%%%%%%%%%%%%%%%%%%%%%%%%%%%%%%

Now we analyze the behavior of the effective mass under changes of the relevant parameters, for the system with compactified spatial dimensions. 
Here we concentrate on the situation of larger values of $ g_{\sigma} $, in which a phase transition of first order takes place. 
 
\begin{figure}[th]
\centering
\includegraphics[{width=8.0cm}]{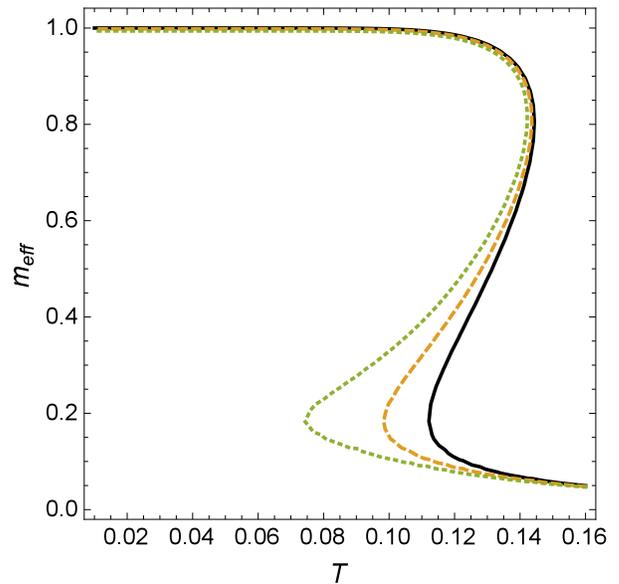}
\caption{Plot of values of $m_{eff}$ that are solutions of the gap equation in Eq. (\ref{eqIIA3}) for $ d=1$ as a function of temperature, at chemical 
equilibrium. We fix $m_{\sigma}=0.53$  and $g_{\sigma}=26.00$. Full, dashed and 
dotted lines represent respectively $L=15.00$, $L=10.00$ and $L=9.00$.}
\label{fig:MassaEfetivaX1}
\end{figure}

In Fig.~\ref{fig:MassaEfetivaX1} is plotted the values of $m_{eff}$ that are solutions of the gap equation in Eq. (\ref{eqIIA3}) as a function of temperature in the case of one compactified spatial dimension (with the scalar density being given by Eq. (\ref{eqIIA5}) for $d=1$). We see that allowed values of effective mass  are 
affected by the presence the of boundary; the range of temperature where occurs the mixed phase is spread out as the length of compactified coordinate decreases. This result suggests that the symmetric phase is favored as  the size of the system decreases. This will be better described via the plots involving the thermodynamic potential density, which will be done as follows.

\begin{figure}[th]
\centering
\includegraphics[{width=8.0cm}]{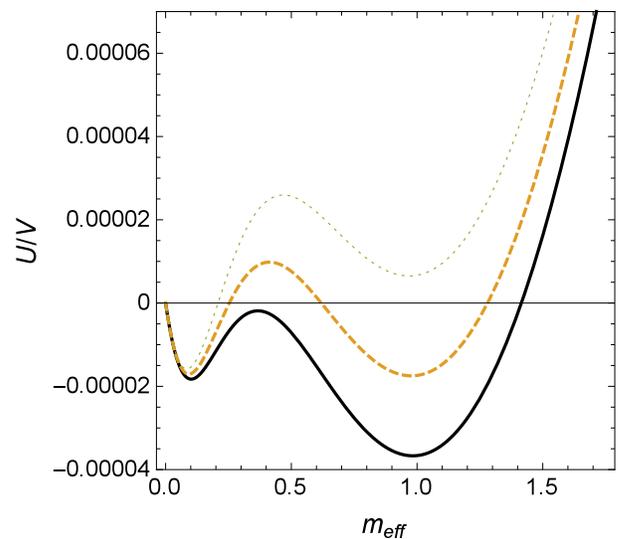}
\caption{Plot of normalized thermodynamic potential density in Eq.~(\ref{norm_pot}) for $d=1$ as a function of effective mass and 
at chemical equilibrium. We fix $m_{\sigma}=0.53$, $g_{\sigma}=26.00$ and $L=15.00$. Full, dashed and dotted lines represent the cases for $T=0.1220$, $T=0.1254$ and $T=0.1293$, respectively.}
\label{fig:PotencialX20}
\end{figure}

\begin{figure}[th]
\centering
\includegraphics[{width=8.0cm}]{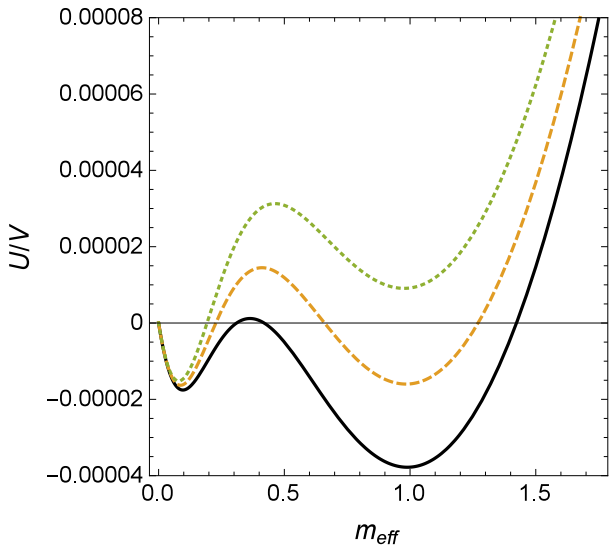}
\caption{Same as in Fig.~\ref{fig:PotencialX20}, but with $L=10.00$. 
Full, dashed and dotted lines represent the cases for $T=0.1150$, $T=0.1200$ and $T=0.1250$, respectively.}
\label{fig:PotencialX12}
\end{figure}

\begin{figure}[th]
\centering
\includegraphics[{width=8.0cm}]{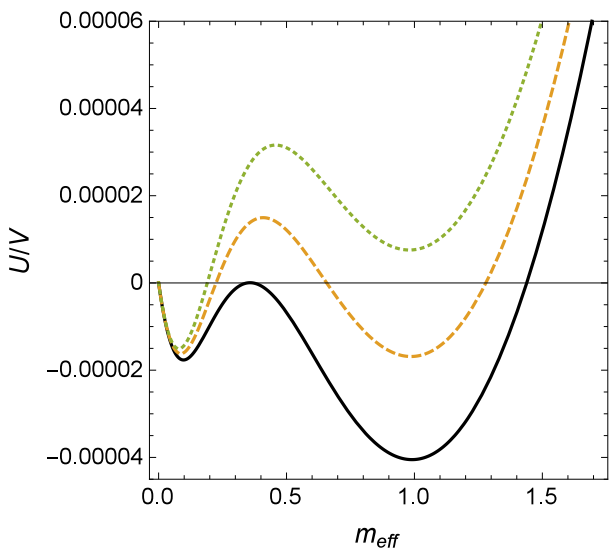}
\caption{Same as in Fig.~\ref{fig:PotencialX20}, but with $L=9.00$. 
Full, dashed and dotted lines represent the cases for $T=0.1050$, $T=0.1127$ and $T=0.1190$,  respectively.}
\label{fig:PotencialX10}
\end{figure}

In Figs.~\ref{fig:PotencialX20}-\ref{fig:PotencialX10} are plotted the normalized thermodynamic potential density defined in Eq. (\ref{norm_pot}) for $d=1$ as a function of effective mass, taking different values of temperature, but with each plot at a given value of the size of the compactified coordinate. These three plots manifest the following nature of the transition: at smaller temperatures the global minimum is at a greater value of $m_{eff}$; the increasing of $T$ makes
the second local minimum at a smaller value of $m_{eff}$
overcome the first one, becoming the absolute minimum; for higher temperatures,  the absolute minimum goes slightly to zero. B

Besides, from  Figs.~\ref{fig:PotencialX20}-\ref{fig:PotencialX10} it can be seen that the decreasing of the size $L$ inhibits the broken phase, in agreement with the analysis concerning Fig. \ref{fig:MassaEfetivaX1}. In other words, smaller values of $L$ induce a first order phase transition at lower critical temperatures.

\begin{figure}[th]
\centering
\includegraphics[{width=8.0cm}]{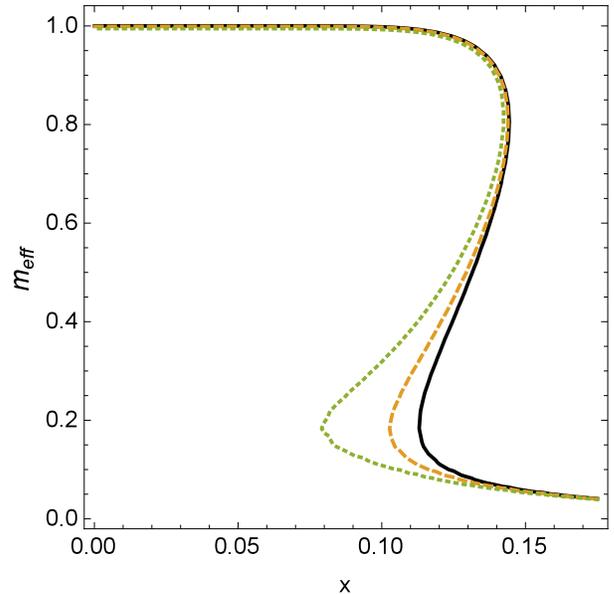}
\caption{Plot of values of $m_{eff}$ that are solutions of the gap equation in Eq. (\ref{eqIIA3}) for $ d=1$ as a function of inverse of length ($x=1/L$), at chemical equilibrium. We fix $m_{\sigma}=0.53$  and $g_{\sigma}=26.00$. Solid, dashed and dotted lines represent respectively $T=0.050$, $T=0.095$ and $T=0.110$.}
\label{fig:MassaEfetivaX2}
\end{figure}

Also, we plot in Fig.~\ref{fig:MassaEfetivaX2} 
the values of effective mass that are solutions of the gap equation in Eq. (\ref{eqIIA3}) as a function of inverse of lentgh $(x=1/L)$, in the case of one compactified spatial dimension (with the scalar density being given by Eq. (\ref{eqIIA5}) for $d=1$). We notice that $m_{eff}$ remains without change at greater values of $L$, where the fluctuations due to size effects are not relevant. However, we remark that the decreasing of the size of the system induces a sudden drop in $m_{eff}$. Namely, there is a critical length $L_c$ of the compactified dimension at which a discontinuous phase transition occurs.  Besides, lower values of  $L_c$ are induced for smaller temperatures.

\begin{figure}[th]
\centering
\includegraphics[{width=8.0cm}]{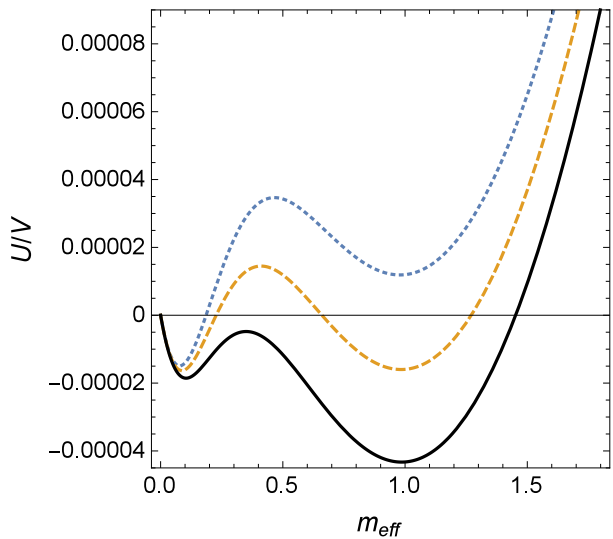}
\caption{Plot of normalized thermodynamic potential density in Eq.~(\ref{norm_pot}) for $d=1$ as a function of effective mass and 
at chemical equilibrium. We fix $m_{\sigma}=0.53$, $g_{\sigma}=26.00$ and $T=0.120$. Full, dashed and dotted lines represent the cases for $L=13.00$, $L=10.00$ and $L=9.00$, respectively.}
\label{fig:PotencialL1L2L3}
\end{figure}

We continue our discussion with Fig.~\ref{fig:PotencialL1L2L3}, in which is plotted the normalized thermodynamic potential density defined in Eq. (\ref{norm_pot}) for $d=1$ as a function of effective mass, taking different values of $L$ and at a fixed temperature.
It can be seen that the global minimum of the system is discontinuously driven  from the regime of greater values of $m_{eff}$ to smaller ones, and it goes slightly to zero as $L$ diminishes.

\begin{figure}[th]
\centering
\includegraphics[{width=8.0cm}]{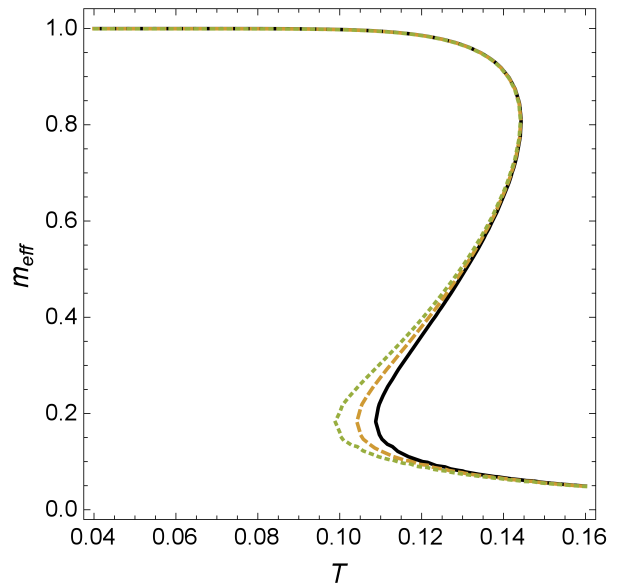}
\caption{Plot of values of $m_{eff}$ that are solutions of the gap equation in Eq. (\ref{eqIIA3}) as a function of temperature, at chemical equilibrium. We fix $m_{\sigma}=0.53$, $g_{\sigma}=26.00$ and $L=12.00$. Solid, dashed and dotted lines represent respectively the case of $d=1$, $d=2$ and $d=3$ compactified spatial dimensions.}
\label{fig:MassaEfetivad123}
\end{figure}

We conclude this investigation with the dependence of the phase structure on  the number of compactified spatial dimensions. 
Then, in Fig.~\ref{fig:MassaEfetivad123} is plotted  
the values of effective mass that are solutions of the gap equation as a function of temperature, for the three situations of compactifed spatial dimensions under study with same length $L$, which is kept fixed. 
Precisely, in this plot is shown the solutions of Eq. (\ref{eqIIA3}), with the scalar density being given by Eq. (\ref{eqIIB4}) for $d=1$, and analogue expressions for $d=2, 3$, setting $L_i = L$ in each case. 
We remark that the range of temperature in which the mixed phase occurs
is spread out as the number of compactified dimensions grows, with the symmetric phase being favored for bigger values of $d$. 
That is, one of the consequences of increasing of the number of compactified spatial dimensions is to cause the decreasing of the temperature at which transition occurs.

\begin{figure}[th]
\centering
\includegraphics[{width=8.0cm}]{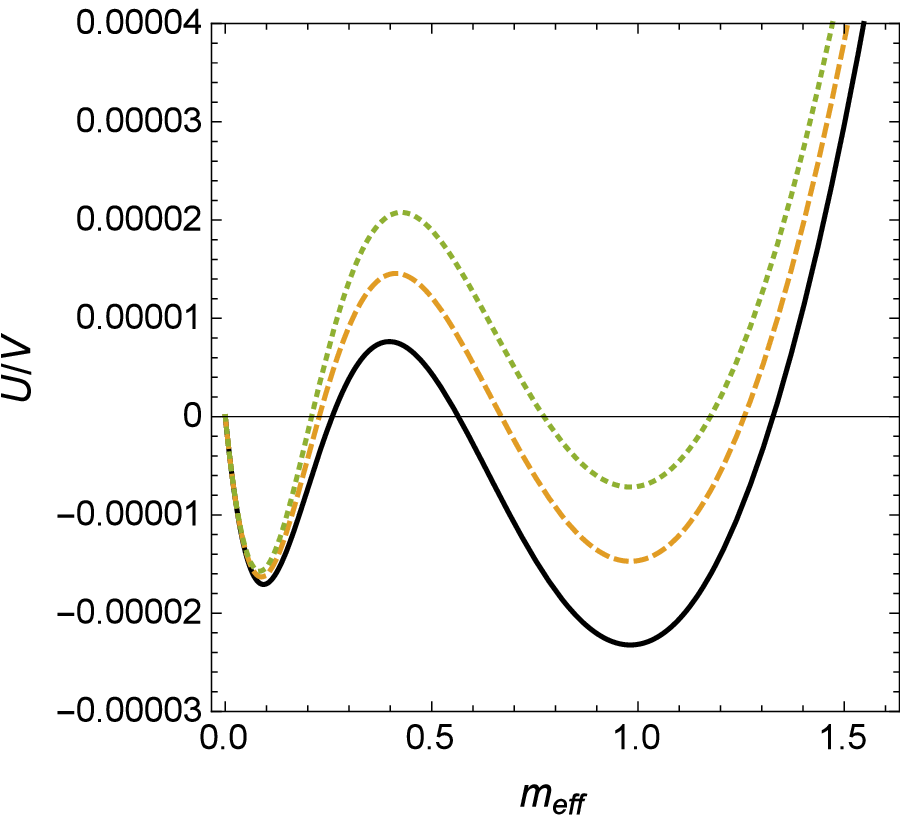}
\caption{Plot of normalized thermodynamic potential density in Eq.~(\ref{norm_pot}) as a function of effective mass and 
at chemical equilibrium. We fix $m_{\sigma}=0.53$, $g_{\sigma}=26.00$, $L=12.00$ and $T=0.123$. Dotted, dashed and solid lines represent the cases for $d=3$, $d=2$ and $d=1$, respectively.}
\label{fig:Potenciald123}
\end{figure}

Once more, the analysis is completed with the normalized thermodynamic potential density defined in Eq. (\ref{norm_pot}) for three situations of compactifed spatial dimensions with same length $L$, as a function of effective mass, at fixed values of temperature and length $L$. This plot is presented in  Fig.~\ref{fig:Potenciald123}
It can be seen that the system is driven from the broken to the disordered phase, as the number of compactified spatial coordinates $d$ increases. 
In other words, our findings suggest that the presence of more boundaries disfavors the maintenance of long-range correlations, making the suppression of the ordered phase.

%%%%%%%%%%%%%%%%%%%%%%%%%%%%%%%%%%%%%%%%%%%%%%%%%%%%%%%%%%%%%%%%%%%%%%%%%%%%%%%%%%%%%%%%
%%%%%%%%%%%%%%%%%%%%%%%%%%%%%%%%%%%%%%%%%%%%%%%%%%%%%%%%%%%%%%%%%%%%%%%%%%%%%%%%%%%%%%%%

\section{Concluding Remarks}\label{Concluding}
%%%%%%%%%%%%%%%%%%%%%%%%%%%%%%%%%%%%%%%%%%%%%%%%%%%%%%%%%%%%%%%%%%%%%%%%%%%%%%%%%%%%%%%%
%%%%%%%%%%%%%%%%%%%%%%%%%%%%%%%%%%%%%%%%%%%%%%%%%%%%%%%%%%%%%%%%%%%%%%%%%%%%%%%%%%%%%%%%

In this work we have investigated the finite-size effects on the phase structure of Walecka model within the framework of generalized Zeta-function, focusing on the influence of temperature as well as the number and length $L_i=L$ of compactified spatial dimensions. As pointed in previous papers, the nature of the phase transition strongly depends on the strength of the coupling constant $ g_{\sigma} $. Here we have concentrated on the situation of larger values of $ g_{\sigma} $, in which a phase transition of first order takes place. 
 The main results obtained are summarized as follows.

We have seen that allowed values of effective mass are 
affected by the presence of boundaries; the range of temperature where occurs the mixed phase is spread as the length of compactified coordinates decreases. This result suggests that the symmetric phase is favored as the size of the system diminishes. In this sense, the decreasing of the size of the system induces a sudden drop in $m_{eff}$. It is suggested a critical length $L_c$ of the compactified dimensions at which a discontinuous phase transition occurs, and lower values of $L_c$ are induced for smaller temperatures.

Besides, it has been remarked that the thermodynamic behavior of the system depends on the number of compactified spatial dimensions $d$. The symmetric phase is favored for bigger values of $d$, with decreasing of the critical temperature. 
The presence of more boundaries tends to inhibit the broken phase. 
In other words, our findings suggest that the presence of more boundaries disfavors the maintenance of long-range correlations, inhibiting of the broken phase. 

Finally, we remark that the findings outlined above may give us insights about relativistic systems that can be interpreted as a fermion gas in a hot medium confined in a reservoir. Further studies will be done in order to apply the present approach to a specific physical system, as the nuclear matter.  Moreover, the present analysis  can be considered as a starting point for further work in which the quantum corrections might be considered.

%%%%%%%%%%%%%%%%%%
%%%%%%%%%%%%%%%%%%
%\section{Acknowledgements}
%{CAPES}
\acknowledgements

The authors would like to thank the Brazilian funding agencies CNPq and CAPES for financial support.
%\vfill \eject
%%%%%%%%%%%%%%%%%%
%%%%%%%%%%%%%%%%%%

%\vfill \eject%
%\newpage%

%%%%%%%%%%%%%%%%%%
%%%%%%%%%%%%%%%%%%

\end{document}